\documentclass[a4paper,12pt]{article} 
\usepackage{graphicx} 
\usepackage{graphics} 

\newcommand{\dfr}{d\raise0.3ex\hbox{\kern-0.5ex\char"013 }} 

\def\l{\left}
\def\r{\right}
\def\be{\begin{equation}}
\def\ee{\end{equation}}
\def\d{\partial}

\begin{document}

\title{Numerical simulation of  a macroscopic quantum-like experiment: oscillating wave packet }
\author{L. Nottale \& Th. Lehner\\{\small CNRS, LUTH, Observatoire de Paris-Meudon,} \\{\small F-92195 Meudon Cedex, France}}
\maketitle

\abstract{\parbox[c]{13cm}{We simulate the transformation of a classical fluid into a quantum-like (super)-fluid by the application of a generalized quantum potential through a retro-active loop. This numerical experiment is exemplified in the case of a non-spreading oscillating wave packet in a harmonic potential. We find signatures of a quantum-like behavior which are stable against various perturbations.}

%*****************
\section{Introduction}
%*****************
One of us has recently proposed a new general concept  of macroscopic quantum-type laboratory experiments  \cite{LN06}. It consists of applying, through a real time retroactive loop, a generalized quantum potential on a classical system. Indeed, one can show that the system of equations (Euler equation and continuity equation) that describes a fluid in irrotational motion subjected to such a generalized quantum potential, that reads $Q=-2{\cal D}^2 {\Delta \sqrt{\rho}}/{\sqrt{\rho}}$ in terms of the density $\rho$, is equivalent to a generalized Schr\"odinger equation.  In this derivation, the quantum potential is no longer founded on the quantum Planck's constant $\hbar$, but on a new constant ${\cal D}$ which can take any macroscopic value. While it would be impossible with present days technology to simulate standard quantum effects by this method because of the smallness of ${\cal D}$, which is given by ${\cal D}=\hbar/2m$ in standard quantum mechanics, the use of a macroscopic value for this constant nevertheless preserves some of the properties of a quantum-like system. Namely, its density distribution is given by the square of the modulus of a complex function which is solution of a Schr\"odinger equation. Therefore such a system is expected to exhibit some quantum-type, superfluid-like macroscopic properties (though certainly not every aspects of a genuine quantum system). 

In the present paper we validate this concept by numerical simulations of a fluid subjected to such a generalized quantum force, as an anticipation of a future real laboratory experiment. The example chosen for this first attempt is the appearance of a non-spreading quantum-like oscillating wave packet in a compressible fluid  (e.g., a plasma) subjected to an attractive harmonic oscillator potential.

%*********************************
\section{Theoretical background}
%*********************************
We consider a classical macroscopic compressible fluid described by the Euler and the continuity equations:
\begin{equation}
\label{AA1}
 \l(\frac{\partial}{\partial t} + V \cdot \nabla\r) V  = -\nabla \phi,
\end{equation}
\begin{equation}
\label{AA2}
\frac{\partial \rho}{\partial t} + {\rm div}(\rho V) = 0,
\end{equation}
where $\phi$ is an exterior scalar potential. We assume as a first step that the pressure term is negligible and  that the fluid motion is potential, i.e.,
\be
V=\nabla S.
\ee
We now assume that we apply to the fluid (using density measurements and a retroaction loop) a varying force which is a function of the fluid density in real time, namely, a  ``quantum-like" force $F_Q$ deriving from the potential
\be
Q=-2{\cal D}^2 \frac{\Delta \sqrt{\rho}}{\sqrt{\rho}}.
\ee
This potential is a generalisation of the standard quantum potential \cite{Bohm}, since here the constant $\cal D$ can have any value, while in standard quantum mechanics it is restricted to the only value $ {\cal D}=\hbar/2m$. As recalled in what follows, this generalization still allows to recover a Schr\"odinger-like equation.

The Euler and continuity system becomes
\begin{equation}
\label{BBB1}
 \l(\frac{\partial}{\partial t} + V \cdot \nabla\r) V  = -\nabla \l({\phi}-2{\cal D}^2 \frac{\Delta \sqrt{\rho}}{\sqrt{\rho}}\r),
\end{equation}
\begin{equation}
\label{BBB2}
\frac{\partial \rho}{\partial t} + {\rm div}(\rho V) = 0,
\end{equation}
The system of equations (\ref{BBB1},\ref{BBB2}) can then be integrated under the form of a generalized Schr\"odinger equation. 

Indeed, equation~(\ref{BBB1}) takes the successive forms
\be
\frac{\d}{\d t}(\nabla S) +\frac{1}{2} \nabla (\nabla S)^2+ \nabla \l({\phi}-2{\cal D}^2 \frac{\Delta \sqrt{\rho}}{\sqrt{\rho}}\r)=0,
\ee
\be
\nabla \l( \frac{\d S}{\d t} +\frac{1}{2}(\nabla S)^2+ {\phi}-2{\cal D}^2 \frac{\Delta \sqrt{\rho}}{\sqrt{\rho}}\r)=0,
\ee
which can be integrated as
\be
 \frac{\d S}{\d t} +\frac{1}{2}(\nabla S)^2+ {\phi}+K-2{\cal D}^2 \frac{\Delta \sqrt{\rho}}{\sqrt{\rho}}=0,
 \ee
where $K$ is a constant that can be renormalized by a redefinition of the potential energy $\phi$. Let us now combine this equation with the continuity equation as follows:
\be
\label{CCC}
\l[-\frac{1}{2} \sqrt{\rho}\l( \frac{\d S}{\d t} +\frac{1}{2}(\nabla S)^2+ {\phi}-2{\cal D}^2 \frac{\Delta \sqrt{\rho}}{\sqrt{\rho}}\r) + i \frac{\cal D}{2\sqrt{\rho}}\l( \frac{\partial \rho}{\partial t} + {\rm div}(\rho \nabla S)  \r) \r] e^{i S/2{\cal D}}=0.
\ee
Finally we set
\be
\label{psi}
\psi=\sqrt{{\rho}} \times e^{{i S}/{2 {\cal D}}},
\ee
and  the equation~(\ref{CCC}) is strictly identical to the following generalized Schr\"odinger equation
\be
{\cal D}^2 \Delta \psi + i {\cal D} \frac{\partial}{\partial t} \psi - \frac{\phi}{2}\psi = 0,
\ee
as can be checked by replacing in it $\psi$ by its expression (\ref{psi}). Recall that such an equation has also been directly obtained, in terms of a density of probability instead of a density of matter, as the integral of the equations of geodesics in a nondifferentiable space-time  \cite{LN93,CN04}. Given the linearity of the equation obtained, one can normalize the modulus of  $\psi$ by replacing the matter density $\rho$ by a probability density  $P=\rho/M$, where $M$ is the total mass of the fluid in the volume considered: this will be equivalent.

The solutions $\psi= |\psi| \times \exp(i \theta)$ of this equation directly provide the density and the velocity field of the fluid at every point, namely
\be
V= 2 {\cal D}\, \nabla \theta, \;\;\; \rho=M |\psi|^2.
\ee
Its imaginary part and its real part amount, respectively, to the continuity equation, and to the energy equation that writes:
\be
E=- \frac{\d S}{\d t} =\frac{1}{2}V^2+ {\phi}-2{\cal D}^2 \frac{\Delta \sqrt{\rho}}{\sqrt{\rho}}.
\ee

The above transformation from the fluid mechanics-like equations to the Schr\"odinger-type equation is similar to a Madelung transformation \cite{Madelung}, but it is here performed in the reversed way and generalized to a constant different from $\hbar/2m$. 

It could be therefore possible by this method to simulate a ``Schr\"odingerian" system, e.g., a partly quantum-like superfluid system coming under two of the axioms of quantum mechanics, namely, it is described by a wave function $\psi$ such that $ \rho \propto |\psi|^2$, which is solution of a Schr\"odinger-type equation.

%****************************************************
\section{Application to the oscillating wave packet}
%****************************************************
As an example of application and as a preparation for a laboratory experiment, let us consider the simplified case of one-dimensional fluid motion in an external harmonic oscillator potential $\phi=(1/2) \omega^2 x^2$. This system is described by the two following equations:
\be
\label{DDD1}
\frac{\partial V}{\partial t} =-V \frac{\partial V}{\partial x}  -\omega^2 x +2{\cal D}^2 \frac{\partial}{\partial x} \l( \frac{\partial^2 \sqrt{\rho}/\partial x^2}{\sqrt{\rho}}\r),
\ee
\be
\label{DDD2}
\frac{\partial \ln \rho}{\partial t} =-\frac{\partial V}{\partial x}-V\,  \frac{\d \ln \rho}{\d x}.
\ee
Here we have written the continuity equation in terms of $\ln \rho$, which will be useful in the numerical simulations that follow. These two equations are equivalent to the one-dimensional generalized Schr\"odinger equation:
\be
{\cal D}^2 \, \frac{\partial^2 \psi}{\d x^2} + i {\cal D}\,  \frac{\partial}{\partial t} \psi - \frac{1}{4} \omega^2 x^2 \psi = 0.
\ee
It is well known that it is possible to find a solution of this equation in the form of a wave packet whose center of gravity oscillates with the period of the classical motion and which shows no spreading with time \cite{Schrodinger26,Schiff68,Landau3}. Assuming that the maximal amount by which the center of gravity is displaced is $a$, the wave function reads in this case
\be
\psi=\l({\frac{\omega}{2 \pi{\cal D}}}\r)^\frac{1}{4}  \; 
e^{-\frac{\omega}{4 {\cal D}}(x-a \cos \omega t )^2}
 \times e^{-i\l(   \frac{1}{2} \omega t + \frac{\omega}{2 {\cal D}} a x \sin \omega t -\frac{\omega}{8 {\cal D}}a^2 \sin 2 \omega t      \r)}.
\ee
Therefore the probability density reads
\be
\label{CI}
P=|\psi|^2=\sqrt{\frac{\omega}{2 \pi{\cal D}}} \; e^{-\frac{\omega}{2 {\cal D}}(x-a \cos\omega t)^2}.
\ee
This is an interesting case for a test of a genuine quantum behavior, since it involves a non vanishing phase in an essential way although this is a one-dimensional system.  The velocity field is given by
\be
V=-a \omega \sin \omega t,
\ee
while the expression for the quantum potential  is
\be
Q(x,t)={\cal D} \omega - \frac{1}{2} \omega^2 (x-a \cos\omega t)^2,
\ee
so that  the quantum force writes
\be
F_Q=-\frac{\d Q}{\d x}= \omega^2 (x-a \cos \omega t).
\ee
Therefore the (varying) energy takes the form
\be
E=\frac{1}{2} V^2 + \phi +Q= {\cal D} \omega + a \omega^2 x \cos \omega t- \frac{1}{2} a^2 \omega^2 \cos(2 \omega t).
\ee
When it is applied to the center of the packet $x=a \cos \omega t$, this expression becomes
\be
E_c={\cal D} \omega + \frac{1}{2} a^2 \omega^2.
\ee
We recognize in the second term, as expected, the energy of a classical pendulum. Concerning the first term, since standard quantum mechanics corresponds to the particular choice ${\cal D}= \hbar/2m$ (here with $m=1$), the term ${\cal D} \omega$ is the generalization of the vacuum energy for an harmonic oscillator, $\frac{1}{2} \hbar \omega$. 

Therefore we verify that the application of a quantum potential on the fluid has given to it some new properties of a quantum-like nature, such as a zero-point energy and the conservation of the shape of the wave packet. 

%*********************************************
\section{Proposed laboratory experiment}
%***********************************************
In order to prepare a real laboratory experiment aiming at achieving such a new macroscopic quantum-like (super)fluid, we shall now present the result of numerical simulations of such an experiment. To this purpose these simulations are not based on the Schr\"odinger form of the equations, but instead on the classical Euler + continuity equations and on the application by feedback of a generalized quantum-like force.

The suggested experiment consists of:

(i) measuring with detectors the density at regular time interval $\{  t_n  \}$ on a grid at positions $\{x_j\}$;

(ii) computing  from these measurements the quantum force $(F_Q)_n=2 {\cal D}^2 \nabla(\Delta \sqrt{\rho_n}/\sqrt{\rho_n})$ at each time $t_n$;

(iii) applying the new value of the force to the fluid at each time $t_n$, therefore simulating by such a feedback the presence of a quantum-like potential. 

The advantage of such a proposal is that one is no longer constrained by the standard quantum value ${\cal D}=\hbar/2m$ that fixes the amplitude of the quantum force, and that one can therefore give to it a macroscopic value, vary it, study its transition to zero (quantum to classical transition), etc...

%*****************
\section{Iterative fitting simulation}
%*****************
\label{simul1}
In this first simulation, we assume that the quantum force (which is a third derivative of the density) is not computed directly from the values of the density, but from a polynomial fit of the distribution of $\ln \rho$. In the special case considered here (the oscillating wave packet), we use a Gaussian fit of the density distribution (i.e. a second order polynomial fit to $\ln \rho$), so that we need to know only the mean and dispersion. More generally, one can decompose the distribution of  $\ln \rho (x)$ into its successive moments.  Therefore the density is written as
\be
\rho_n(x) \propto \exp\l[-\frac{1}{2}\l(\frac{x-\bar{x}_{n}}{\sigma_n}\r)^2\r],
\ee
so that, once the mean and dispersion $\bar{x}_{n}$ and ${\sigma_n}$ at time $t_n$ are computed, the quantum force to be applied at each step ($n$) writes:
\be
(F_Q)_n(x)= \frac{{\cal D}^2(x-\bar{x}_{n})}{\sigma_n^4}.
\ee

\subsection{Numerical simulation}
\label{sim1}
%***********************************
Our numerical simulation is performed by a simple Mathematica program which reproduces the steps of the real experiment, namely, at each time step $t_n$: 

(i) We compute the mean and the dispersion of positions $x$ according to the density distribution:
\be
\bar{x}= \sum_j \rho(x_j) x_j / \sum_j \rho(x_j),
\ee
\be
\sigma^2= \sum_j \rho(x_j) (x_j -\bar{x})^2/ \sum_j \rho(x_j).
\ee

(ii) The force $F_Q$ to be added then writes in terms of these quantities
\be
(F_Q)_n(x)= \frac{{\cal D}^2(x-\bar{x}_{n})}{\sigma_n^4 \; \delta x^3},
\ee
where $ \delta x$ is the grid interval and intervenes here because we use finite differences.

(iii) We compute the logarithm of the density $\ln \rho$  and the velocity $V$ at next time step $t_{n+1}$ by transforming equations (\ref{DDD1}, \ref{DDD2}) into centered finite-difference equations (Forward Time Centered Space, FTCS scheme) using the Lax-Friedrichs method \cite{NumRec}, namely,
\be
\ln \rho_j^{n+1}=\frac{\ln \rho_{j+1}^n+\ln \rho_{j-1}^n}{2}  -\frac{\delta t}{2\, \delta x} \l\{ (V_{j+1}^n -V_{j-1}^n)+V_j^n  \, (\ln \rho_{j+1}^n -\ln \rho_{j-1}^n)  \r\},
\ee
\be
V_j^{n+1}=\frac{V_{j+1}^n+ V_{j-1}^n}{2}+ \delta t   \l(    -V_j^n \;  \frac{V_{j+1}^n  -V_{j-1}^n}{2 \,  \delta x} + F_j^n+({F_Q})_j^n   \r).
\label{EulerNum}
\ee
The lower index ($j$) is for space $x$ and the upper one ($n$) is for time $t$;  $\delta t$ is the time step and $F(x)= - \omega^2 x$ is the external harmonic oscillator force. In the above Lax method, the terms $\ln \rho_j^n$ and $V_j^n $ are replaced by their space average, which has the advantage to stabilize the FTCS scheme. 

The  initial conditions are given by the density distribution (Eq.~\ref{CI}) for $t=0$.

Although this is a simple scheme (we have not attempted at this stage to better control numerical error diffusion), it has given very encouraging results, since it has reproduced on several periods the expected motion of the quantum oscillating wave packet (see Figure~\ref{fig1}). 

%%%%%%%%%
\begin{figure}[!ht]
\begin{center}
\includegraphics[width=8cm]{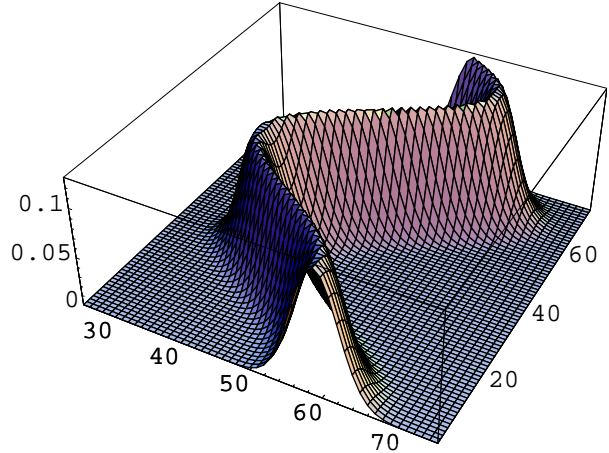}
\caption{\footnotesize{Result of the numerical integration of a Euler + continuity one dimensional system with generalized quantum potential for the oscillating wave packet in an harmonic oscillator field. The quantum force applied on the fluid is calculated from a gaussian fit of the density distribution. The figure gives the density distribution obtained in function of position (space grid from 25 to 75) and time (time steps from 1 to 75, i.e., 1.2 period).}}
\label{fig1}
\end{center}
\end{figure}
%%%%%%%%

%*****************
\subsection{Perturbation of initial conditions}
%*****************
One of the possible shortcomings in the passage from the simulation to a real experiment may come from fluctuations in the initial conditions. Indeed, in the previous simulations, we have taken as initial density distribution that of the exact quantum wave packet. In order to be closer to a real experimental situation, we have therefore performed a new simulation similar to that of  Sec.~\ref{sim1}, but with an initial density distribution that is perturbed with respect to the Gaussian solution (Eq.~\ref{CI}): we have multiplied its values $\rho(x_j)$ at each point $\{x_j\}$ of the space grid by $\exp(\alpha)$, where $\alpha$ is random in the interval [0,1]. A typical resulting initial density distribution is given in Figure~\ref{fig2}, followed by the distributions obtained on a full period (sub-figures 1 to 12) after application of the generalized quantum force. 

Once again the result obtained is very enrouraging as concerns the possibility of performing a real laboratory experiment, since, despite the initial deformation, the wave packet remains stable during several periods. Moreover, not only the mean and dispersion of the evolving density distribution remain close to the ones expected for the quantum wave packet, but, as can be seen in Figure~\ref{fig2}, the initial perturbations have even been smoothed out during the feedback process.

%%%%%%%%
\begin{figure}[!ht]
\begin{center}
\includegraphics[width=8cm]{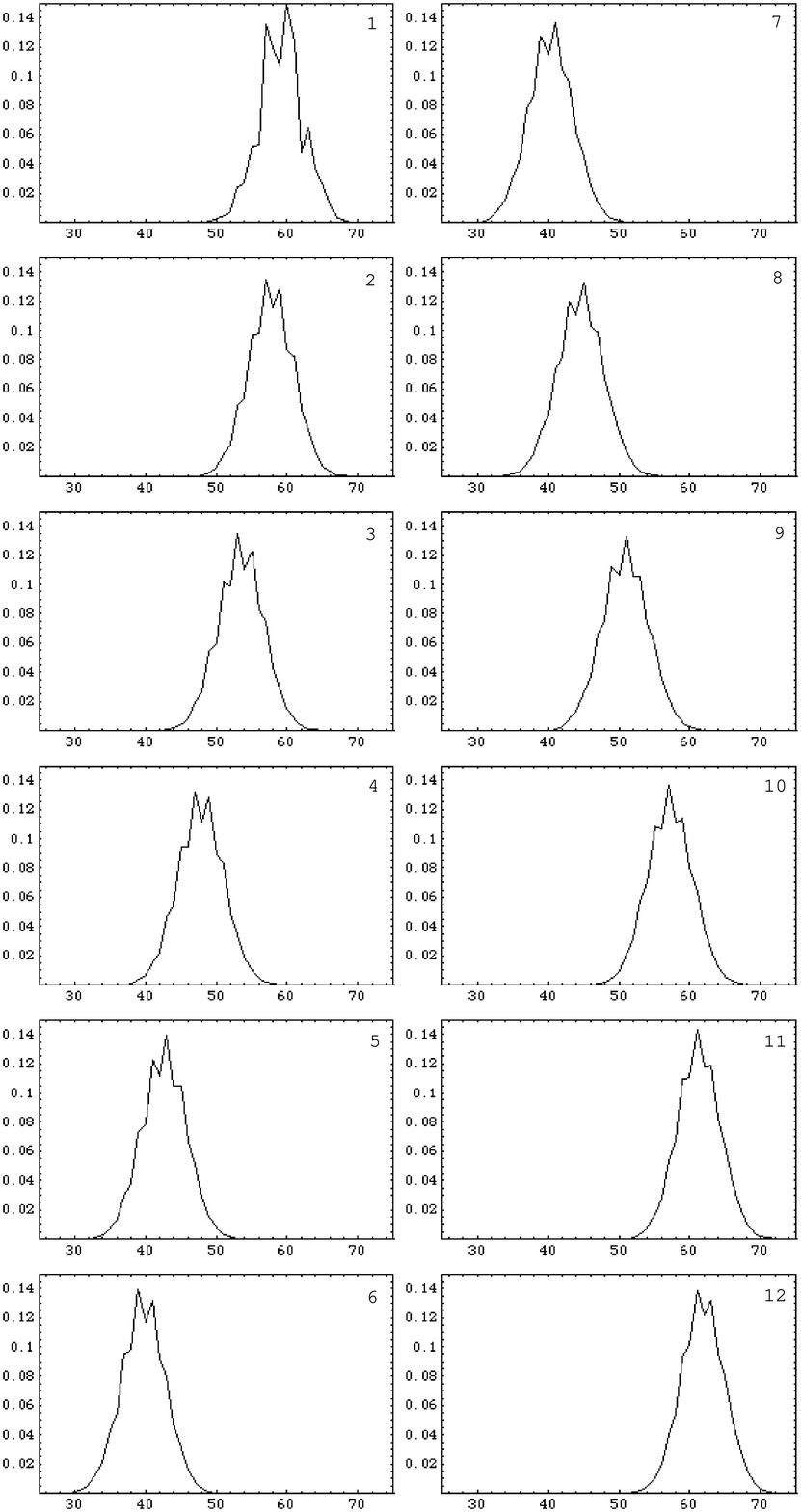}
\caption{\footnotesize{Result of the numerical integration of a Euler + continuity one dimensional system with added quantum potential for the oscillating wave packet. The conditions are the same as in Figure \ref{fig1}, except for the addition of a perturbation on the initial density distribution (left top figure). The quantum force applied on the fluid is calculated from a Gaussian fit of the density distribution. The successive figures give the density distribution obtained in function of position (space grid from 25 to 75) and time (64 time steps corresponding to one period, among which twelve of them, equally distributed, are shown). }}
\label{fig2}
\end{center}
\end{figure}

%*****************
\subsection{General account of uncertainties}
%****************
This encouraging result leads us to attempt a numerical simulation under far more difficult conditions: in order to simulate the various uncertainties and errors that may occur in a real experiment, in particular as concerns the density measurement, the application of the force, and  physical effects not accounted in the simulation such as pressure (see below), vorticity, etc..., we have now added a fluctuation at each step of the retroactive loop. Namely, at each time step $t_n$, we have multiplied the density $\rho(x_j)$ at each point $\{x_j\}$ of the space grid by $\exp(\alpha)$, where $\alpha$ is random in the interval [0,1].

As can be seen in Figure~\ref{fig3}, despite the large errors added, the numerical simulation shows an oscillating wave packet which, despite its large fluctuations, keeps its coherence. In particular, it keeps the values of the mean and dispersion (to about 5 percent) expected for the quantum solution on about 1/3 of period before the end of the simulation due to numerical errors.

%%%%%%%%
\begin{figure}[!ht]
\begin{center}
\includegraphics[width=8cm]{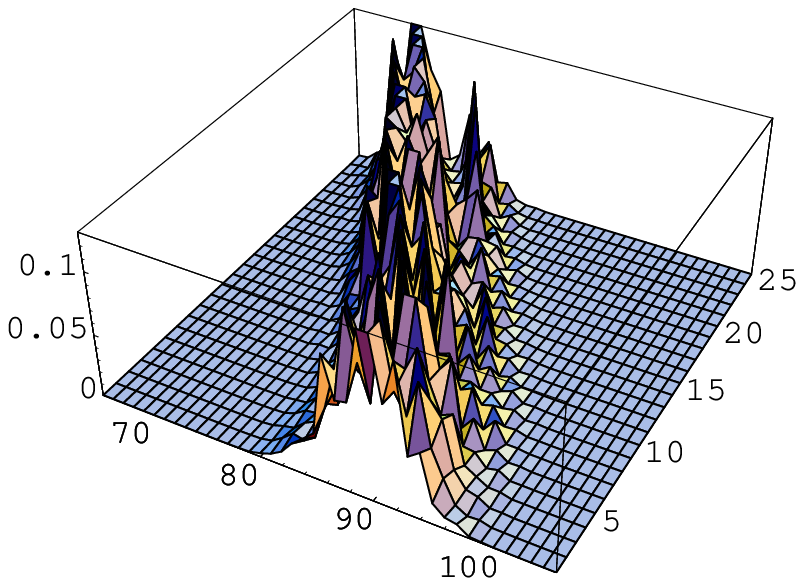}
\caption{\footnotesize{Result of the numerical integration of a Euler + continuity one dimensional system with added generalized quantum potential for the oscillating wave packet. The conditions are the same as in Figure \ref{fig1}, except for the addition of a perturbation on the density distribution at each time step of the simulation. The quantum force applied on the fluid is calculated from a Gaussian fit of the density distribution. The successive figures give the density distribution obtained in function of position (space grid from 65 to 105) and time (time steps 1 to 25, which corresponds to almost half a period). }}
\label{fig3}
\end{center}
\end{figure}
%%%%%%%%%

%*****************
\subsection{Account of pressure}
%****************
The addition of a pressure term in the initial Euler equation still allows one to obtain a Schr\"odinger-like equation in the general case when $\nabla p/\rho$ is a gradient, i.e., $\nabla p/\rho=\nabla w$. This is the case of an isentropic fluid, and, more generally, of every cases when there is an univocal link between pressure and density, e.g., a state equation \cite{Landau6}. The Euler equation with quantum potential and external potential reads
\begin{equation}
 \l(\frac{\partial}{\partial t} + V \cdot \nabla\r) V  = -\nabla \l({\phi}+w-2{\cal D}^2 \frac{\Delta \sqrt{\rho}}{\sqrt{\rho}}\r),
\end{equation}
and it can therefore, in combination with the continuity equation, be integrated in terms of a Schr\"odinger-like equation, 
\be
{\cal D}^2 \Delta \psi + i {\cal D} \frac{\partial}{\partial t} \psi - \frac{\phi+w}{2} \, \psi = 0.
\ee
Now the pressure term needs to be specified through a state equation, which can be chosen as taking the general form $p=k_p \rho^{\gamma}$. The special case $\gamma=1$ can be recovered and its amplitude established by taking the acoustic limit $p=p_0+p'$, $\rho=\rho_0+\rho'$ and $p'=c_s^2 \rho'$, where $c_s$ is the sound velocity in the fluid. Therefore one obtains a linear relation $p=a+ c_s^2 \rho$, so that the pressure term in the Euler equation finally reads $\nabla p/ \rho=k_p \nabla \ln \rho$, while $w=k_p \ln \rho=k_p \ln |\psi|^2$, with $k_p=c_s^2$. This means that the integrated equation is now a nonlinear Schr\"odinger equation, 
\be
{\cal D}^2 \Delta \psi + i {\cal D} \frac{\partial}{\partial t} \psi -k_p \ln |\psi| \; \psi =\frac{1}{2} \phi  \; \psi.
\ee
In the highly compressible case the dominant pressure term  is rather  $\propto \rho^2$, and the $ \ln |\psi|$ term is replaced by $ |\psi|^2$ in the non-linear Schr\"odinger equation (see e.g. \cite{Nore}).

%%%%%%%%
\begin{figure}[!ht]
\begin{center}
\includegraphics[width=8cm]{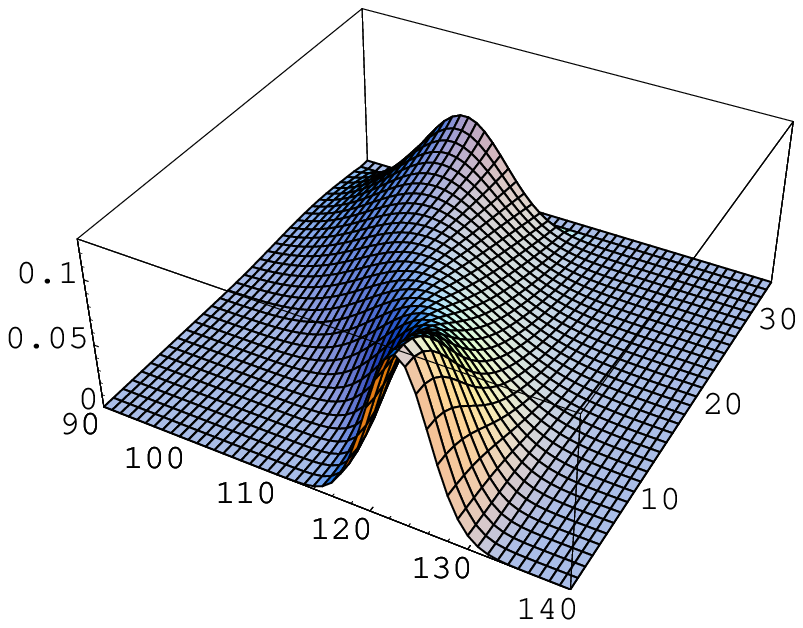}
\caption{\footnotesize{Result of the numerical integration of a Euler and continuity one-dimensional system of equations with added generalized quantum potential and account of a pressure term, for the oscillating wave packet. The quantum force applied on the fluid is calculated from a Gaussian fit of the density distribution. The figure gives the probability density in function of position (space grid from 90 to 140) and time (time steps from 1 to 32).  In this simulation (near half a period), the amplitude of the pressure term is $k_p=5$.}}
\label{fig4}
\end{center}
\end{figure}
%%%%%%%%%

%%%%%%%%
\begin{figure}[!ht]
\begin{center}
\includegraphics[width=8cm]{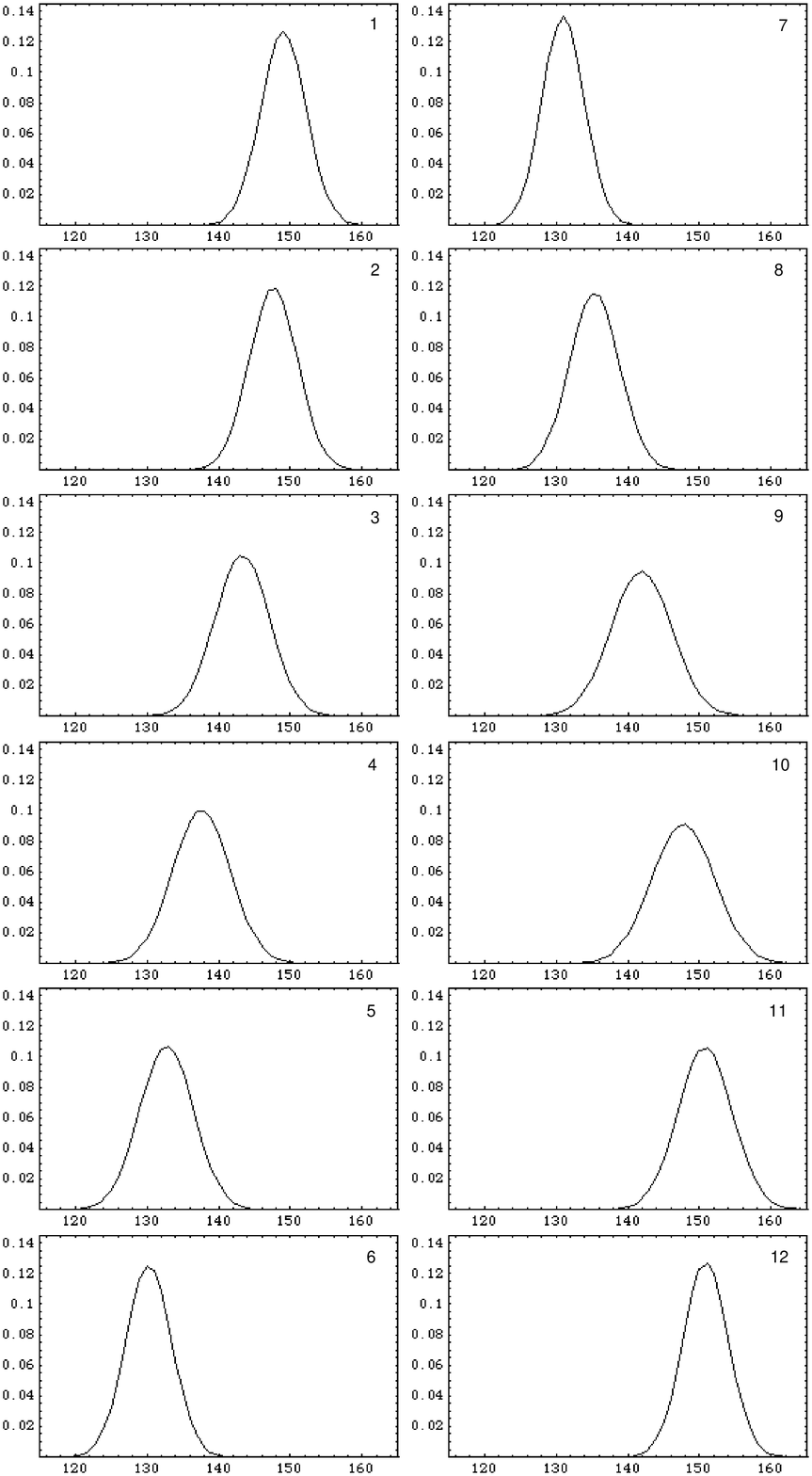}
\caption{\footnotesize{Result of the numerical integration of a Euler and continuity one-dimensional system of equations with added generalized quantum potential and account of a pressure term, for the oscillating wave packet. The quantum force applied on the fluid is calculated from a Gaussian fit of the density distribution.  The figure gives the density distribution in function of the position (space grid form 120 to 160), for 12 equal time steps which cover a full period. In this simulation, the amplitude of the pressure term is $k_p=1$. One sees that the effect of pressure amounts to an oscillating spreading of the wave packet, which nearly recovers its shape after half a period.}}
\label{fig5}
\end{center}
\end{figure}

The numerical integration is now performed by generalizing Eq.~(\ref{EulerNum}) as
\be
V_j^{n+1}=\frac{V_{j+1}^n+ V_{j-1}^n}{2}+ \delta t   \l(    -V_j^n \;  \frac{V_{j+1}^n  -V_{j-1}^n}{2 \,  \delta x} + F_j^n+({F_Q})_j^n - k_p  \frac{\ln \rho_{j+1}^n -\ln \rho_{j-1}^n}{2\delta x} \r).
\ee
The result is given in Figures \ref{fig4} and \ref{fig5} for two different values of the pressure amplitude $k_p$. One finds that the addition of pressure leads to an oscillatory slight spreading of the wave packet, but that its main superfluid-like features are preserved, since it nearly recovers its shape after half a period.

%******************************************
\section{Full finite differences simulation}
%*****************************************
The success of this first simple simulation leads us to attempt a more direct feedback in which the quantum force is computed by finite differences from the values of the density itself (while in the previous simulation we used an intermediate polynomial fit from which the force was analytically derived).

To this purpose, we use a form of the generalized quantum potential and of the generalized quantum force according to which they can be expressed in terms of only $\nabla \ln P$ (or equivalently $\nabla \ln \rho$). Setting
\be
H=\nabla \ln P,
\ee
we find:
\be
Q=-{\cal D}^2   \l( \nabla . H+ \frac{1}{2} H^2  \r),
\ee
\be
F_Q=-\nabla Q={\cal D}^2   \l[  \Delta H+  (H.\nabla)H  \r].
\label{FQ}
\ee
In one dimension it reads
\be
F_Q={\cal D}^2 \l( \frac{\d^3 \ln P}{\d x^3} +  \frac{\d^2 \ln P}{\d x^2} \: \frac{\d \ln P}{\d x} \r) .
\ee
The numerical integration proceeds following the same lines as in the previous simulation, except for the first steps aiming at computing $F_Q$, which are replaced by a finite difference calculation according to equation (\ref{FQ}). Such a way to compute the force $F_Q$ to be applied on the fluid is therefore directly similar to its calculation in a real laboratory experiment from digitalized measurements of the density by a grid of detectors. Namely, we calculate successively, for all values of the position index $j$,
\be
H_j^n=\frac{\ln \rho_{j+1}^n-\ln \rho_{j-1}^n}{2 \, \delta x},
\ee
then similar relations for positions $x_{j-1}$,  $x_{j+1}$, $x_{j-2}$ and  $x_{j+2}$, then
\be
Q_{j-1}^n=-{\cal D}^2 \l\{ \frac{H_{j}^n - H_{j-2}^n} {2 \, \delta x}+ \frac{1}{2} \l(H_{j-1}^n\r)^2 \r\},
\ee
then a similar relation for $Q(x_{j+1},t_n)$, and finally
\be
(F_Q)_{j}^n= \frac{Q_{j-1}^n - Q_{j+1}^n}{2 \, \delta x}.
\ee

The calculation of $\ln \rho$ (from the continuity equation) and of $V$ (from the Euler equation) are the same as previously. We have attempted to use other more precise formulas for the calculation of the second and third order derivatives in the expression of $F_Q$: this has led to essentially the same result.

Despite, once again, the roughness of the chosen integration method, the result obtained is satisfactory, since the motion of a quantum non-spreading oscillating wave packet has been reproduced on about 1/4 of period before divergence due to the effect of computing errors (Figure \ref{fig6}). This result has been obtained without using the Schr\"odinger equation, but instead an apparently  ``classical" hydrodynamic Euler/continuity system with an externally applied generalized quantum  potential.

%%%%%%%%
\begin{figure}[!ht]
\begin{center}
\includegraphics[width=7cm]{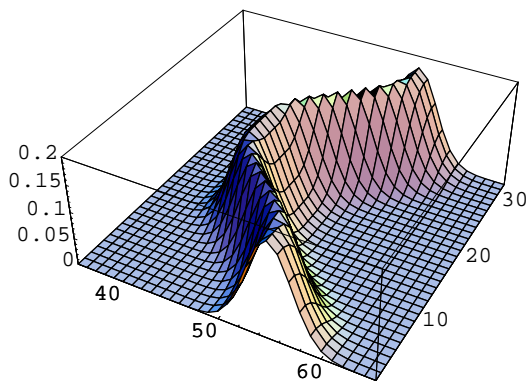}
\caption{\footnotesize{Result of the numerical integration of a Euler + continuity one dimensional system with generalized quantum potential for the oscillating wave packet in an harmonic oscillator field. The quantum force applied on the fluid is directly calculated from the values of the density by finite differences. The density distribution obtained in function of position (space grid from 35 to 65) on about 1/4 of period has been used to reconstruct a full period (32 time steps).}}
\label{fig6}
\end{center}
\end{figure}
%%%%%%%%

Adding a pressure term yields a similar result  (i.e., reproduction of the motion of the wave packet on about 1/4 of period before divergence due to the effect of computing errors) which confirms the result obtained with the Gauss fitting method, namely, a partial oscillating spreading of the wave packet (Figure \ref{fig7}).

%%%%%%%%
\begin{figure}[!ht]
\begin{center}
\includegraphics[width=7cm]{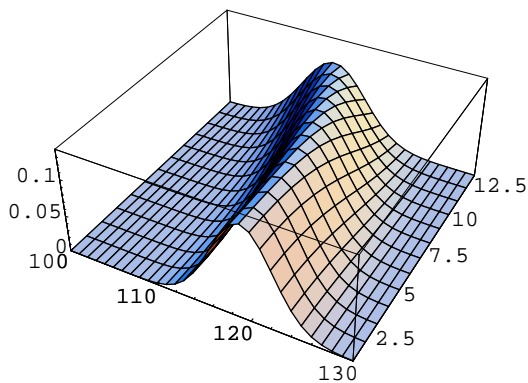}
\caption{\footnotesize{Result of the numerical integration of a Euler + continuity one dimensional system with generalized quantum potential for the oscillating wave packet in an harmonic oscillator field and account of pressure (1/4 of period before stop due to computing errors). The quantum force applied on the fluid is here directly calculated from the values of the density by finite differences. The figure gives the probability density in function of space (grid from 100 to 130) and time (time steps from 1 to 13). A pressure term has been added ($k_p=1$, whose effect is a slight oscillatory spreading of the wave packet. }}
\label{fig7}
\end{center}
\end{figure}
%%%%%%%%

%%%%%%%%
%************************************
\section{Discussion and conclusion}
%*************************************
These preliminary simulations were intended to yield a first validation of the concept of a new kind of quantum-like macroscopic experiments based on the application to a classical system of a generalized quantum force through a retroaction loop \cite{LN06}. They have given a positive results, since the expected quantum-type stable structure (here a non-spreading or slightly spreading oscillating wave packet) has been obtained during a reasonably long time of integration. These results, obtained by a rather rough integration method, are very encouraging since they give the hope that a real laboratory experiment should be possible to achieve.

In the hydrodynamic case considered in this work, possible shortcomings are to be considered, such as the effects of finite compressibility, of vorticity, of viscosity at small scales, of density detector uncertainties, of the minimal time interval needed to perform the loop for the calculation and the application of the quantum force in a real experiment, etc... 

We have attempted here to have a first account of these uncertainties by taking a pressure term into account, by adding large random fluctuations in the initial conditions, then by adding large fluctuations at each time steps of the simulation. The results obtained were again encouraging, since, despite the pressure term and the large fluctuations, the overall coherence of the wave packet and its period were preserved.  We shall in forthcoming works attempt to take into account these effects in more complete numerical simulations with improved integration schemes, apply the same general concept to other types of systems, then lead a real hydrodynamic laboratory experiment \cite{LehnerNottale06}. 

Provided such an actual experiment succeeds, it could lead to many new applications in several domains: didactic ones (teaching of quantum mechanics), laboratory physics (macroscopic models of quantum systems, simulations of atomic and molecular systems, study of the quantum to classical transition, laboratory astrophysics \cite{LN97,NSL00}, models of biological-like systems \cite{LN04}), technology (development of a  new devices having some quantum-like properties and behavior), self-organization (plama confinement, control of turbulence ?, etc..).
\\

Acknowledgements. The authors gratefully acknowledge very fruitful discussions with Dr. L. Di Menza.

%*****************
\end{document}